\documentclass[a4paper]{jpconf}

\usepackage{graphicx}
\usepackage{icsghead}
\usepackage{amssymb}

\usepackage{subfigure}

\usepackage{hyperref}

\begin{document}
\title{
Simplifying Generalized Belief Propagation on Redundant Region Graphs
}

\author{Chuang Wang and Hai-Jun Zhou}

\address{
State Key Laboratory of Theoretical Physics,
Institute of Theoretical Physics,
Chinese Academy of Sciences,
Zhong-Guan-Cun East Road 55, Beijing 100190, China
}

\ead{chuangphys@gmail.com, zhouhj@itp.ac.cn}

\begin{abstract}
    The cluster variation method has been developed into a general theoretical framework
    for treating short-range correlations in many-body systems after it was first 
    proposed by Kikuchi in 1951.
    On the numerical side, a message-passing approach called
    generalized belief propagation (GBP) was proposed by
    Yedidia, Freeman and Weiss about a decade ago
    as a way of computing the minimal value of the
    cluster variational free energy and
    the marginal distributions of clusters of variables.
    However the GBP equations are often redundant, and it is
    quite a non-trivial task to make the GBP iteration converges to a fixed point.
    These drawbacks hinder the application of the GBP approach to
    finite-dimensional frustrated and disordered systems.

    In this work we report an alternative and simple derivation of the GBP
    equations starting from the partition function expression.
    Based on this derivation we propose a natural and systematic way of removing
    the redundance of the GBP equations. We apply the simplified generalized
    belief propagation (SGBP) equations to the two-dimensional and
    the three-dimensional ferromagnetic Ising model and Edwards-Anderson
    spin glass model. The numerical results
    confirm that the SGBP message-passing approach is able to
    achieve satisfactory performance on these model systems. We also suggest that
    a subset of the SGBP equations can be neglected in the numerical iteration process
    without affecting the final results.
\end{abstract}

\section{Introduction}

Short loops are abundant in finite-dimensional ferromagnetic spin models and
spin glass models. They cause complicated and strong local correlations in
the systems.
The accumulation and propagation of these local correlations then lead to
long-range correlations and the emergence of various collective behaviors.
The cluster variation method (CVM) is a general theoretical framework for
treating local correlations in many-body statistical systems.
The original idea of the
cluster variation method was conceived by Professor Ryoichi Kikuchi
(1919-2003) in $1951$ \cite{Kikuchi-1951}. Since then CVM has
been applied to many different types of systems and has been further
developed and generalized
\cite{An-1988,Morita-etal-1994,Opper-Saad-2001,Pelizzola-2005,Rizzo-etal-2010}.
The basic idea of CVM is to decompose the entropy of the whole system
into the residual entropy contributions of various clusters
of spin variables. After this decomposition, the true entropy of the
system is approximated by the sum of residual entropy contributions from
a properly chosen subset of spin clusters,
see \cite{An-1988} or section~2.2 of
\cite{Zhou-Wang-2012} for a brief introduction.

The celebrated Bethe-Peierls tree approximation
\cite{Bethe-1935,Peierls-1936a,Peierls-1936,Chang-1937}
is an important limiting case of the cluster variation method. This
tree approximation plays a central role in the mean-field
theory of spin glasses \cite{Mezard-etal-1987,Mezard-Montanari-2009},
and it is also underlying the widely used
belief propagation (BP) message-passing algorithm in
information science \cite{Pearl-1988}. For various spin glass problems
defined on finite-connectivity random graphs, due to the absence of
short loops, the Bethe-Peierls approximation or its extended version
(with the possibility of ergodicity breaking in the
configuration space being considered)
can give asymptotically
exact results in the thermodynamic limit (see
\cite{Mezard-Montanari-2009} for a comprehensive review). But
the Bethe-Peierls approximation is inadequate in treating strong local
correlations and therefore it performs poorly on finite-dimensional
spin glass systems.

Considering more local correlations
beyond the level of Bethe-Peierls
approximation is conceptually easy within the CVM framework,
but for strongly disordered and frustrated spin glass systems
this task is  practically quite challenging. There are two major issues:
(a) How to construct a suitable variational marginal probability distribution
for each chosen cluster of spin variables? (b) How to efficiently minimize
the total Kikuchi cluster variational free energy, a
complicated function of a large set of parameters?
Yedidia and co-authors proposed in
\cite{Yedidia-Freeman-Weiss-2005} a particular way of constructing
cluster marginal probability distribution functions, and then
they suggested a generalized belief propagation (GBP) message-passing
approach to minimize the resulting Kikuchi variational free energy.
This GBP message-passing approach outperforms the BP message-passing approach
considerably in terms of numerical precision, but its widespread
applications on finite-dimensional spin glass systems are still hindered
by two important drawbacks: first, the GBP equations are often redundant;
and second, it is quite a non-trivial task to make the GBP iteration
converges to a fixed point.

In this work we understand the cluster marginal probability distribution
functions of \cite{Yedidia-Freeman-Weiss-2005} from the viewpoint of
the equilibrium partition function, and give an alternative and
simple derivation of the GBP equations.
Based on this derivation we propose a natural and systematic way of removing
the redundance of the GBP equations. The resulting simplified generalized
belief propagation (SGBP) equations are much more convenient for
numerical implementation compared with the original GBP
equations.
We also point out that, for a given system, a subset of the SGBP equations 
can be safely ignored in the numerical iteration process.
 As redundance is minimized, the
iteration process based on SGBP is much more easier to
converge to a fixed point.
We demonstrate the good  performance of the SGBP equations by
applying these equations to the two-dimensional (2D) and
the three-dimensional (3D) ferromagnetic Ising model and Edwards-Anderson
spin glass model.

The next section defines the general model system. The GBP equations are
derived in section~\ref{sec:pfrg} starting from the partition function.
These equations are then simplified in section~\ref{sec:sgbp}.
Some numerical results obtained by the SGBP equations are discussed
in section~\ref{sec:EA}.
We conclude this work in section~\ref{sec:conclusion}.

\section{The general model system}

We first define the general model system  and briefly
describe the region graph concept.
A convenient expression for the partition function of the system is given.

\subsection{Energy}

Consider a system with $N$ vertices and $M$
two- or many-body   interactions among the vertices.
The indices of the vertices are denoted as $i, j, k, \ldots \in [1, N]$ and
those of the   interaction terms as $a, b, c, \ldots \in [1, M]$.
Each vertex $i$ has a state $x_i$ which, for notational simplicity,
is assumed to be a discrete scalar variable.
A microscopic configuration of the system is denoted as $\underline{x}$, with
$\underline{x} \equiv \{x_1, \ldots, x_i, \ldots, x_N\}$.
The energy function has the following general form
\begin{equation}
\label{eq:totalenergy}
H(\underline{x}) = \sum\limits_{i=1}^{N} E_i(x_i)
 + \sum\limits_{a=1}^{M} E_a(\underline{x}_{\partial a}) \; .
\end{equation}
In the above expression, $E_i(x_i)$ and $E_a(\underline{x}_{\partial a})$ are,
respectively, the self-energy of vertex $i$ and the energy of
  interaction $a$; $\partial a$ denotes the
set of vertices involved in interaction $a$,
and $\underline{x}_{\partial a} \equiv \{x_i | i \in \partial a\}$ is
a microscopic sub-configuration for this particular set of vertices.

To give a concrete example of the general model system (\ref{eq:totalenergy}),
let us mention the Edwards-Anderson (EA) spin glass system
on a finite-dimensional regular lattice
\cite{Edwards-Anderson-1975}. The vertices are then the lattice sites,
each of them having a binary spin state
($x_i \in \{-1, +1\}$). There is a spin coupling interaction
between each pair of nearest neighboring lattice sites.
 The energy of a given spin configuration is
\begin{equation}
\label{eq:EAh}
H(\underline{x})
= - \sum\limits_{i=1}^{N} h_i^0 x_i - \sum\limits_{(i,j)} J_{i j}
x_i x_j \; ,
\end{equation}
where $h_i^0$ is the external field on vertex (lattice site)
$i$, and $(i,j)$ denotes a pair of nearest-neighboring vertices $i$ and $j$,
with $J_{i j}$ being the coupling constant between them.

\begin{figure}
  \centering
  \subfigure[]{
    \label{fig:2Drg:a}
    \includegraphics[width=0.3\textwidth]{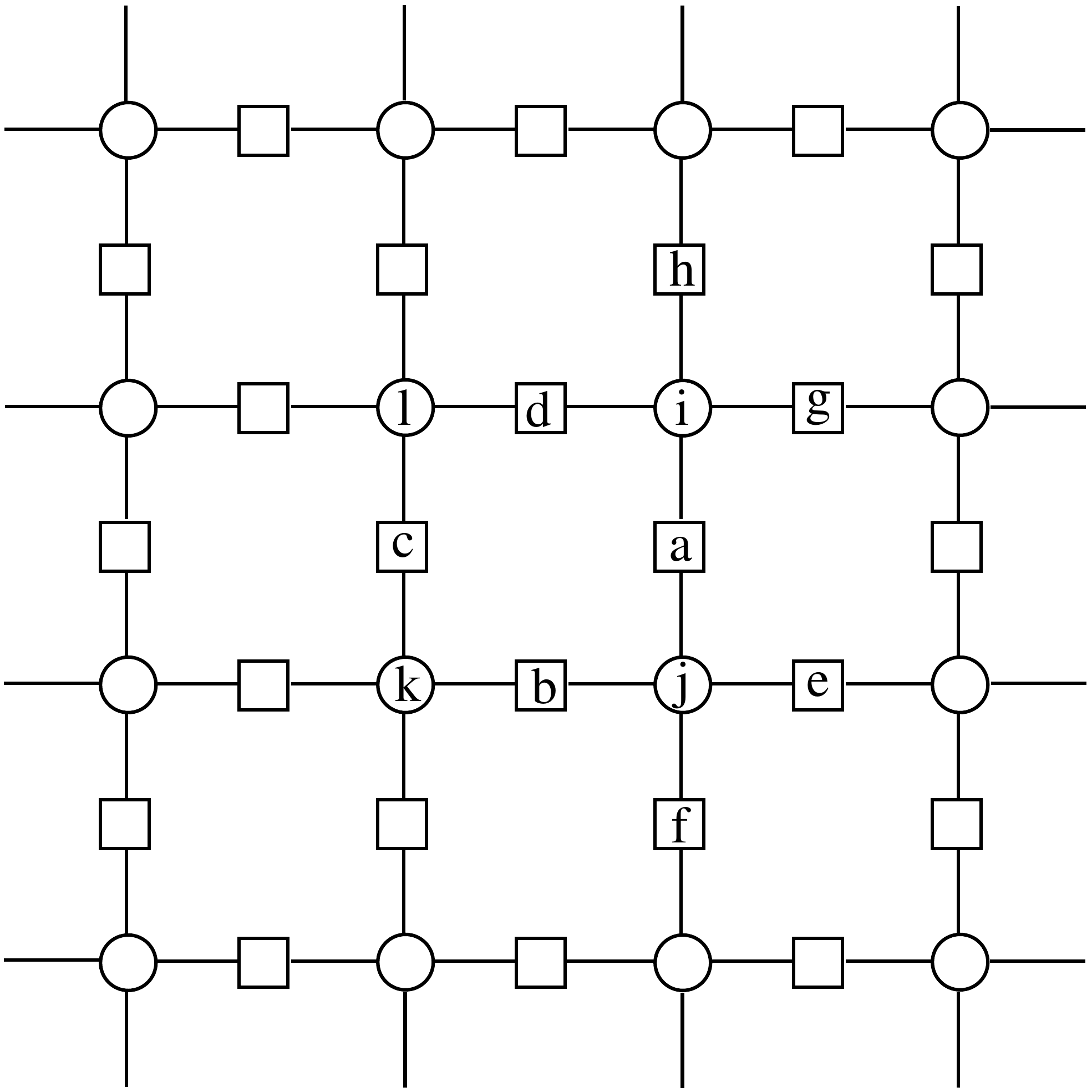}
  }
  \hspace{1cm}
  \subfigure[]{
    \label{fig:2Drg:b}
    \includegraphics[width=0.3\textwidth]{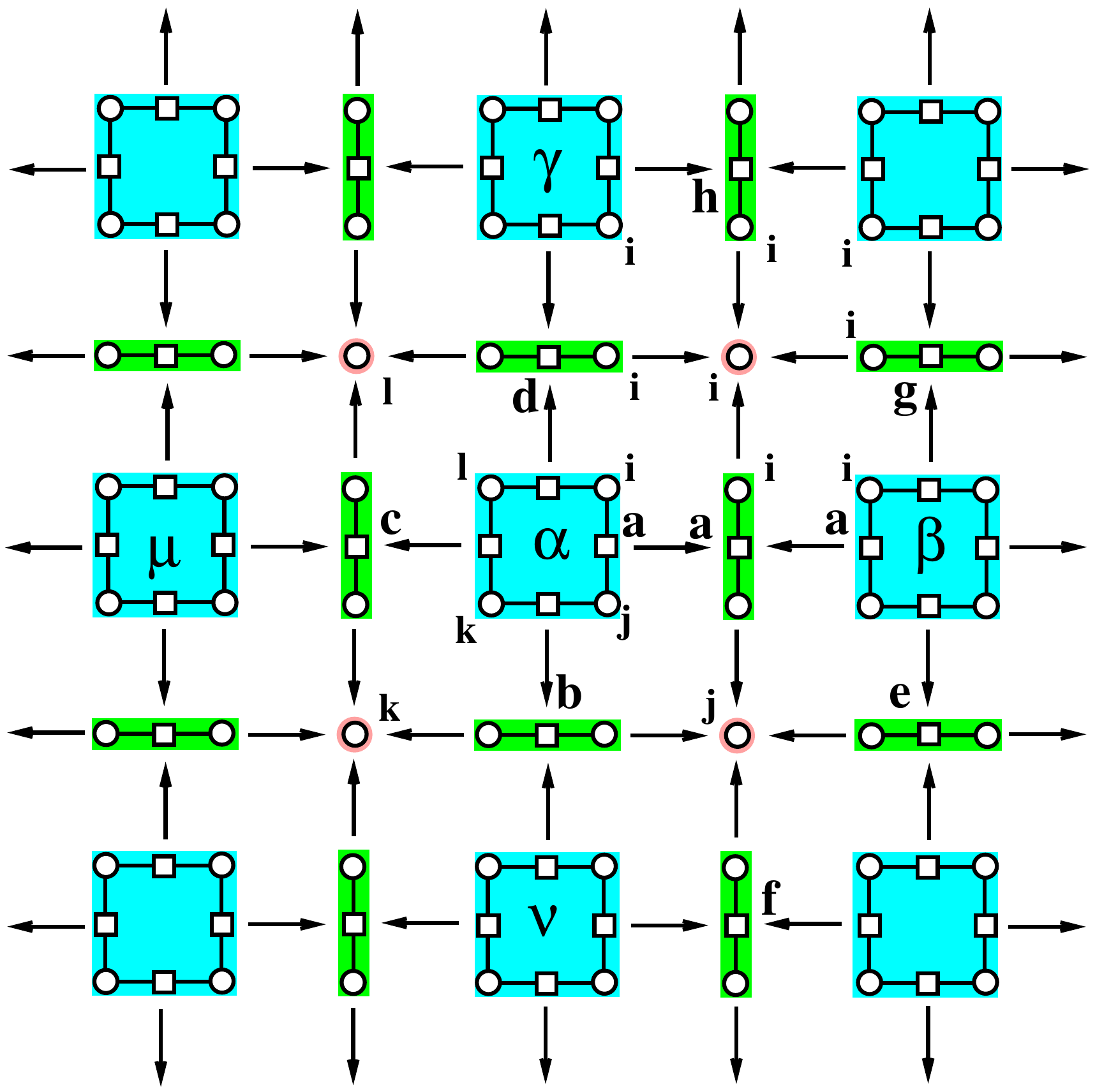}
  }
  \caption{
    (a) Part of the factor graph for the Edwards-Anderson model on
    a square lattice (with periodic boundary condition on each side).
    Each small circle (such as $i$) denotes a vertex (lattice site);
    each small square (such as $a$) denotes a spin coupling interaction. An
    edge $(i, a)$ between a circle $i$ and a square $a$ means that vertex $i$
    participates in interaction $a$.
    (b) Part of a redundant region graph for the factor graph of (a).
    Each square region (highlighted in cyan, such as $\alpha$ and $\beta$)
    contains four vertices and four interactions;
    each rod region (highlighted in green, such as $a$ and $b$) contains two
    vertices and one interaction; and each vertex region (highlighted in
    red, such as $i$ and $j$)
    contains only one vertex. A directed edge points from a parent region to
    a child region. Each vertex appears in $9$ regions (see the example
    of vertex $i$); and each interaction appears in $3$ regions (see the
    example of interaction $a$ between vertex $i$ and $j$). The counting
    number for a square region, a rod region and a vertex region is
    $c=1$, $c=-1$ and $c=1$, respectively.}
  \label{fig:2Drg}
\end{figure}

A conventional graphical representation for
the general model system  (\ref{eq:totalenergy}) is a bipartite graph
of $N$ small circles (representing the vertices) and $M$ small squares
(representing the interactions). Such a bipartite graph is
referred to as a factor graph in the literature \cite{Frey-1998}.
Each edge in the factor graph is between a small circle and a small square.
If and only if a vertex $i$ is
involved in an interaction $a$, then there will be an edge
connecting the corresponding small circle and small square in the factor graph.
As an example, figure~\ref{fig:2Drg:a} shows part of the factor graph
for the EA model (\ref{eq:EAh}) on a periodic square lattice.

\subsection{Partition function and free energy}

The partition function of system (\ref{eq:totalenergy}) has the
sum-product form
\begin{equation}
 \label{eq:partitionfunction}
    Z(\beta) \equiv \sum_{\underline{x}} e^{-\beta H(\underline{x})}
  = \sum\limits_{\underline{x}} \prod\limits_{i=1}^{N} \psi_i(x_i)
 \prod\limits_{a=1}^{M} \psi_a(\underline{x}_{\partial a}) \; .
\end{equation}
The inverse temperature $\beta \equiv \frac{1}{k_B T}$, with $T$ being
the temperature (we shall set Boltzmann's constant $k_B=1$ in the following
discussions).
The functions $\psi_i(x_i)$ and $\psi_a(\underline{x}_{\partial a})$ are,
respectively, the Boltzmann
factor for the self-energy $E_i$ and the interaction energy $E_a$,
\begin{equation}
\psi_i(x_i)  \equiv   e^{-\beta E_i(x_i)} \; , \quad\quad\quad
\psi_a(\underline{x}_{\partial a})
 \equiv  e^{-\beta E_a(\underline{x}_{\partial a})} \; .
\end{equation}
The equilibrium free energy $F(\beta)$ is related with
the partition function $Z(\beta)$ as
\begin{equation}
F(\beta)  \equiv -\frac{1}{\beta} \ln Z(\beta) \; .
\label{eq:freeenergy2}
\end{equation}

Knowing the free energy $F$ as a function of inverse temperature $\beta$
and (if necessary)
other environmental control parameters, we can then calculate all the
other thermodynamic quantities such as the mean energy, the entropy,
the mean value of $x_i$ for each vertex $i$. The free energy and the
partition function therefore have fundamental importance in equilibrium
statistical mechanics.
But exactly computing the partition function (and the
free energy) is an impossible task generically.
Many numerical schemes have been developed over the years to
obtain good approximate values for $Z(\beta)$ \cite{Opper-Saad-2001}.

A well-known theoretical framework for performing approximation
is Kikuchi's cluster variation method
\cite{Kikuchi-1951,An-1988,Morita-etal-1994,Pelizzola-2005}.
However, the variational problem of minimizing
the Kikuchi cluster free energy is rather difficult to solve, especially
for spin glass systems with quenched random parameters.
Yedidia, Freeman, and Weiss \cite{Yedidia-Freeman-Weiss-2005} demonstrated
that the minimal points of the Kikuchi variational free energy correspond
to fixed points of a set of self-consistent generalized belief propagation
(GBP) equations. Minimisation of the Kikuchi cluster free energy was therefore
turned into the problem of constructing a fixed point for the GBP equations,
which can be achieved through an iterative message-passing process on a
region graph (see next subsection).
Unfortunately, the GBP iterative process is still
computationally demanding, especially when the underlying region graph
are redundant.

In this work we will give an alterative and simple derivation of
the GBP equations starting from the partition
function (\ref{eq:partitionfunction}). An advantage of this derivation is that
it suggests a natural way of simplifying the GBP equations on redundant
region graphs.

\subsection{Region graph representation}
\label{subsec:rg}

The basic motivation of the region graph concept is to
facilitate treatments of local correlations through distributing vertices
into different overlapping groups, with each group containing a subset of
vertices that are believed to be strongly correlated
\cite{Kikuchi-1951,An-1988,Yedidia-Freeman-Weiss-2005}.

A region graph $R$ is a graph composed of regions and directed edges between
pairs of regions \cite{Yedidia-Freeman-Weiss-2005}. The regions are
generally denoted by Greek symbols. Each region $\gamma$ contains
a subset of the
vertices and a subset of the self-energies and interaction energies.
If there is a directed edge from a region $\mu$
to another region $\nu$, denoted as $\mu\rightarrow \nu$, then $\nu$ must be
contained in $\mu$ (namely all the vertices and all the energy contained
in $\nu$ must also be contained in $\mu$).
When the edge $\mu\rightarrow \nu$ is present, we say that $\mu$ is a
parent of $\nu$ and $\nu$ a child of $\mu$.
If there is a directed path from a region $\alpha$ to another region
$\gamma$, we say that $\alpha$ is an ancestor
of $\gamma$ and $\gamma$ a descendant of $\alpha$, and denote this
ancestor--descendant
relationship by $\alpha > \gamma$ and $\gamma < \alpha$.
The notation $\alpha \geq \gamma$ ($\alpha \leq \gamma$)
is understood as either $\alpha$ and $\gamma$ are identical to each other
or $\alpha$ is an ancestor (descendant) of $\gamma$.

Figure~\ref{fig:2Drg:b} shows part of a region graph for the EA model of
figure~\ref{fig:2Drg:a}. There are
three types of regions. Each square region contains four vertices and
four interactions, and it is
parent of four rod regions;  each rod region contains two vertices and
one interaction, and it is parent of  two vertex regions; each
vertex region contains a single vertex. In this particular example,
for notational simplicity,
each rod region is
denoted by the index of the single interaction in it, and each vertex region is
denoted by the index of the single vertex in it.

Each region $\gamma$ is assigned a counting number $c_\gamma$ \cite{An-1988},
which  is constructed recursively by
\begin{equation}
 \label{eq:rgcn}
c_\gamma = 1 - \sum\limits_{\{\alpha \; : \; \alpha>\gamma\}} c_\alpha \; .
\end{equation}
If a region $\alpha$ has no directed edges pointing to it, its counting number
is $c_\alpha = 1$. For a region $\mu$ with ancestors,  it is
obvious from the construction (\ref{eq:rgcn}) that
$c_\mu + \sum_{\gamma > \mu} c_\gamma  \equiv 1$.

The vertices and energy terms of system (\ref{eq:totalenergy})
are clustered into various
regions of $R$. This clustering, however, is not exclusive.
A vertex and an energy term may be assigned to more than one region.
To ensure that each
energy term contributes only one Boltzmann factor to the partition function
(\ref{eq:partitionfunction}), the region graph $R$
is required to satisfy the following two constraints
\cite{Yedidia-Freeman-Weiss-2005}:
(1) For any vertex $i$, the induced subgraph $R_i$ (i.e., the region subgraph
formed by all the
regions containing $i$ and all the directed edges between pairs of these
regions) is connected, and the sum of counting numbers within $R_i$ is unity:
\begin{equation}
 \label{eq:cni}
    \sum\limits_{\gamma \in R_i} c_\gamma = 1 \; ;
\end{equation}
and (2) for any interaction $a$, the induced subgraph $R_a$ (i.e., the
region subgraph formed by all the
regions containing $a$ and all the directed edges between pairs of these
regions) is connected, and the
sum of counting numbers within $R_a$ is unity:
\begin{equation}
\label{eq:cna}
\sum\limits_{\gamma \in R_a} c_\gamma = 1 \; .
\end{equation}
Because of (\ref{eq:cni}) and (\ref{eq:cna}),
the partition function can be expressed as
\begin{equation}
\label{eq:Z-exp-1}
    Z(\beta) = \sum\limits_{\underline{x}}
    \prod\limits_{\alpha \in R} \biggl[
    \prod_{i\in \alpha} \psi_i(x_i) \prod_{a\in \alpha}
\psi_a(\underline{x}_{\partial a})
    \biggr]^{c_\alpha} \; .
\end{equation}
In our earlier work \cite{Zhou-etal-2011,Zhou-Wang-2012},
the expression (\ref{eq:Z-exp-1})  was the starting point for
performing partition function expansion on the region graph $R$.

\begin{figure}
    \centering
    \subfigure[]{
    \label{fig:2Dgroup:a}
    \includegraphics[width=0.3\textwidth]{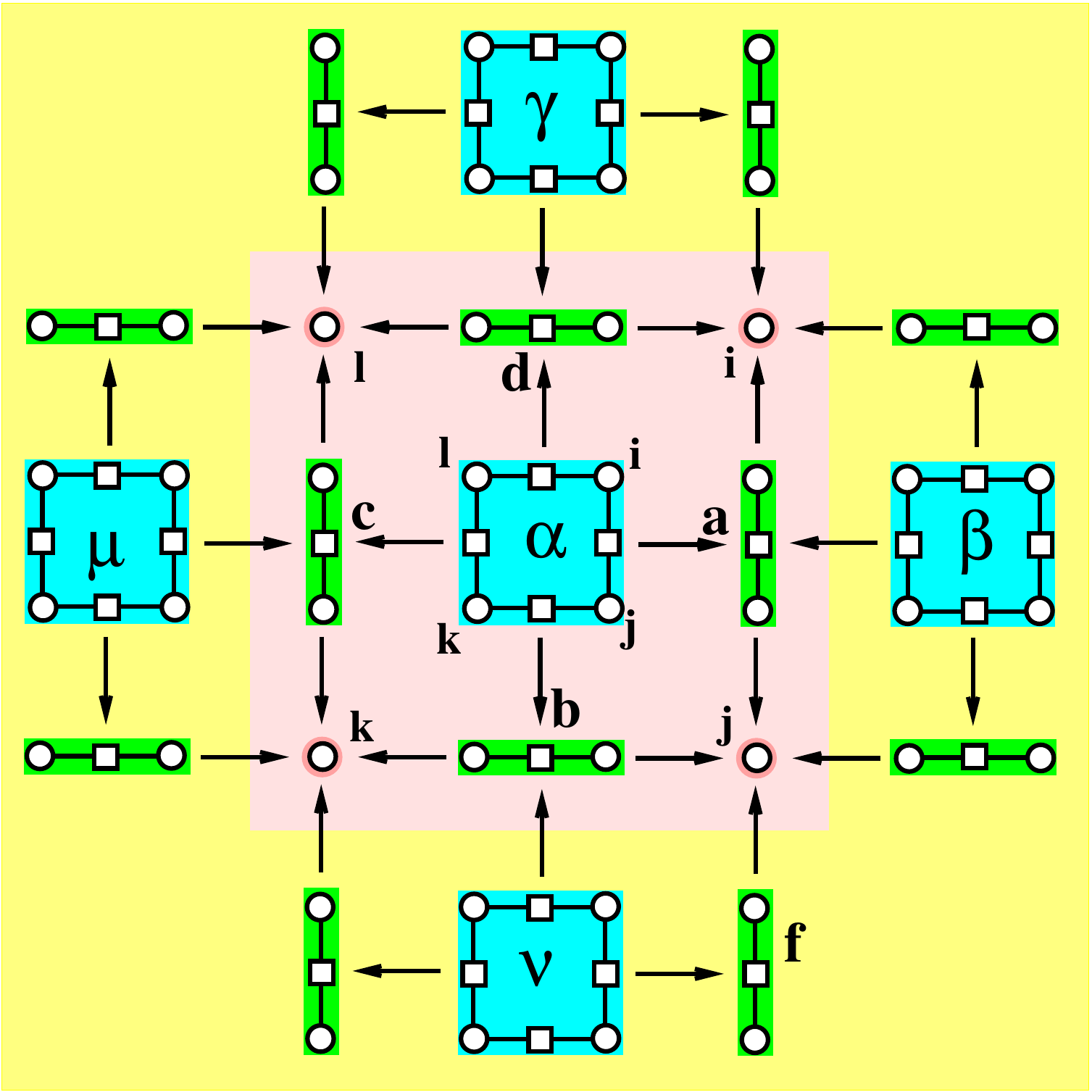}}
    \hspace{1cm}
    \subfigure[]{
    \label{fig:2Dgroup:b}
    \includegraphics[width=0.185\textwidth]{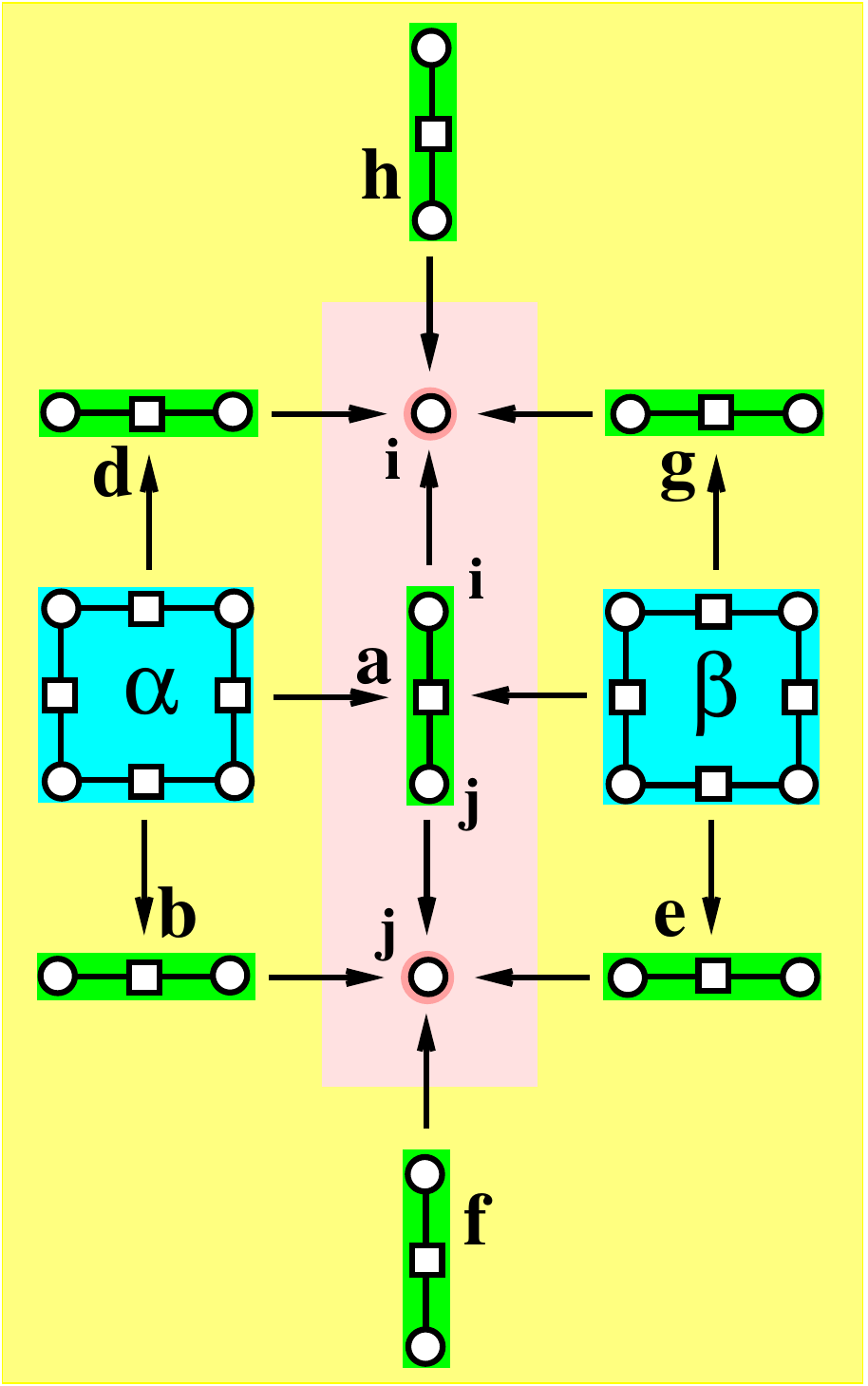}}
    \hspace{1cm}
    \subfigure[]{
    \label{fig:2Dgroup:c}
    \includegraphics[width=0.185\textwidth]{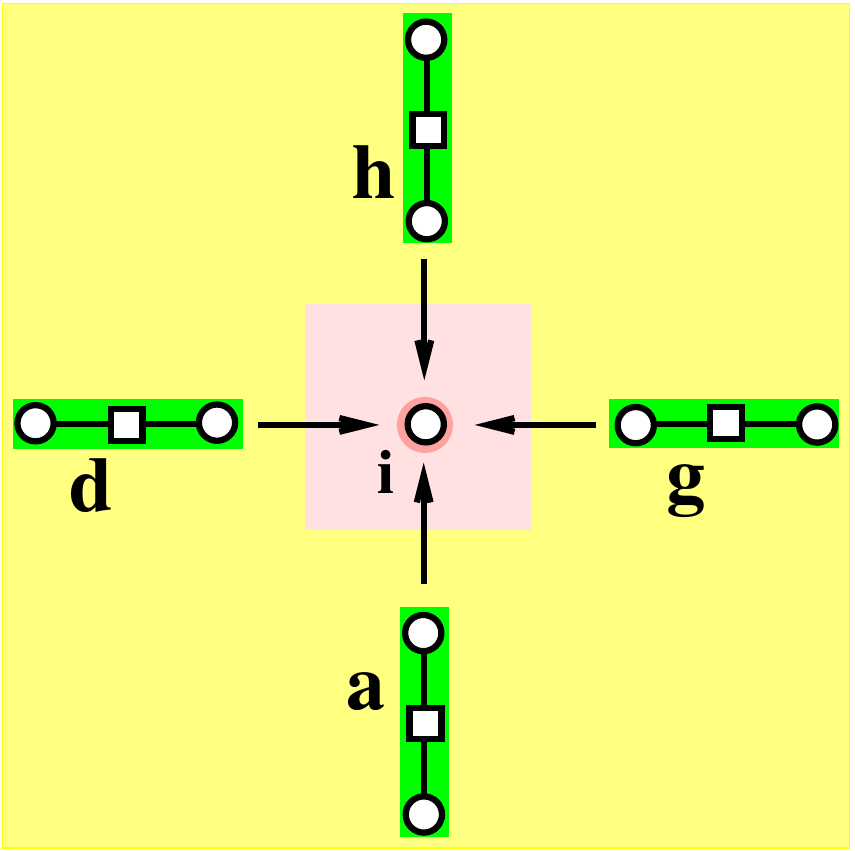}}
    \caption{The interior and boundary for the square region $\alpha$ (a), the
     rod region $a$ (b), and the vertex region $i$ (c) of
    figure~\ref{fig:2Drg:b}. The
     interior set $I_\gamma$ of any region $\gamma$ is marked by a
    pink-colored domain,
     while the boundary set $B_\gamma$ of  region $\gamma$
     is marked by a yellow-colored domain.}
     \label{fig:2Dgroup}
\end{figure}

For each region $\alpha$ of a given region graph $R$,
let us denote by $I_\alpha \equiv \{ \gamma: \gamma \leq \alpha\}$
the set formed by
region $\alpha$ and all its descendants,
by $A_\alpha \equiv \{\gamma : \gamma > \alpha\}$ the set formed by
all the regions ancestral to region $\alpha$,
and by $B_\alpha$ the set that contains all the regions
not belonging to set $I_\alpha$ but parental to at least
one region of set $I_\alpha$ \cite{Zhou-Wang-2012}. We refer to
the set $I_\alpha$ as the interior of
region $\alpha$ and set $B_\alpha$
as the boundary of region $\alpha$. As some examples, we show in
figure~\ref{fig:2Dgroup} the interiors and boundaries of
the square region $\alpha$, the rod region $a$ and the vertex region $i$ of
the region graph of figure~\ref{fig:2Drg:b}.

A region graph $R$ is referred to as \emph{non-redundant} if the region
subgraph $R_i$ induced by any vertex $i$ is a tree (containing no loops),
otherwise $R$ is referred to as \emph{redundant}
\cite{Zhou-etal-2011,Zhou-Wang-2012}.
In a non-redundant region graph there is only one directed path from
a region $\alpha$ to a descendant region $\gamma$, while in a
redundant region graph there may exist multiple directed paths from an
ancestral region $\alpha$ to a descendant region $\gamma$.
The region graph shown in
figure~\ref{fig:2Drg:b} is redundant, since there are two
directed paths ($\alpha\rightarrow a \rightarrow i$ and
$\alpha\rightarrow d \rightarrow i$) from the square region $\alpha$ to
the vertex region $i$.

\section{Generalized belief propagation (GBP)
equations}
\label{sec:pfrg}

We give an alternative derivation of the generalized belief propagation (GBP)
equations \cite{Yedidia-Freeman-Weiss-2005} in this section.
Let us introduce on each directed edge $\mu\rightarrow \nu$ of the
region graph $R$ an arbitrary probability distribution function
$m_{\mu\rightarrow \nu}(\underline{x}_\nu)$, with the
only constraints that this function is positive and
is properly normalized,
$\sum_{\underline{x}_\nu} m_{\mu\rightarrow \nu}(\underline{x}_\nu)
\equiv 1$. The function $m_{\mu\rightarrow \nu}(\underline{x}_\nu)$
is a probability measure on the microscopic
states $\underline{x}_\nu \equiv \{x_i | i\in \nu\}$ of region $\nu$.
This measure is applied on $\nu$ by the parent region $\mu$. The probability
$m_{\mu\rightarrow \nu}(\underline{x}_{\nu})$ can be regarded as a
message from the parent region $\mu$ to the child region $\nu$ concerning
the microscopic sub-configuration $\underline{x}_{\nu}$.

We observe that, for each directed edge $\mu\rightarrow \nu$,
\begin{equation}
\label{eq:c-identity}
\sum\limits_{\{\alpha \ : \ \mu \in B_\alpha, \nu \in I_\alpha\}} c_\alpha
= \sum\limits_{\alpha \geq \nu} c_\alpha -
\sum\limits_{\alpha \geq \mu} c_\alpha = 1-1 = 0 \; .
\end{equation}
This identity guarantees that the
partition function (\ref{eq:Z-exp-1}) can be expressed as
\begin{equation}
Z(\beta)  =
\sum\limits_{\underline{x}}
\prod\limits_{\alpha\in R}
\biggl[
\prod\limits_{i\in \alpha} \psi_i(x_i)
\prod\limits_{a\in \alpha} \psi_a(\underline{x}_{\partial a})
\prod\limits_{\{\mu\rightarrow \nu \
: \
\mu\in B_\alpha, \nu\in I_\alpha
\}} m_{\mu\rightarrow \nu}(\underline{x}_\nu)
\biggr]^{c_\alpha}  \; .
\label{eq:Z-exp-2}
\end{equation}

For each region $\alpha$ let us define a Boltzmann factor $z_\alpha$ as
\begin{equation}
\label{eq:zalpha}
z_\alpha  = \sum\limits_{\underline{x}_\alpha}
\prod\limits_{i\in \alpha} \psi_i(x_i)
\prod\limits_{a\in \alpha} \psi_a(\underline{x}_{\partial a})
\prod\limits_{\{\mu\rightarrow \nu \ : \ \mu \in B_\alpha, \nu \in I_\alpha\}}
m_{\mu\rightarrow \nu}(\underline{x}_\nu) \; .
\end{equation}
Then the partition function can be re-written as
\begin{equation}
\label{eq:Z-exp-3}
Z(\beta) = Z_0 \times
\sum\limits_{\underline{x}} \prod\limits_{\alpha \in R}
\omega_\alpha(\underline{x}_\alpha)^{c_\alpha} \; ,
\end{equation}
where $Z_0 \equiv \prod\limits_{\alpha \in R} z_\alpha^{c_\alpha}$, and
the weight $\omega_\alpha$ is defined as
\begin{equation}
\label{eq:omega}
\omega_\alpha(\underline{x}_\alpha)
  \equiv \frac{1}{z_\alpha} \prod\limits_{i\in \alpha} \psi_i(x_i)
\prod\limits_{a\in \alpha} \psi_a(\underline{x}_{\partial a})
\prod\limits_{\{\mu\rightarrow \nu \ : \ \mu \in B_\alpha, \nu \in I_\alpha\}}
m_{\mu\rightarrow \nu}(\underline{x}_\nu) \; .
\end{equation}
By exploiting the expression (\ref{eq:Z-exp-3}) we can write the free energy
$F(\beta)$ as $F(\beta) = F_0 + \Delta F$, with
\begin{equation}
\label{eq:F0}
F_0 \equiv - \frac{1}{\beta} \ln Z_0 = -\frac{1}{\beta}
\sum\limits_{\alpha \in R}
c_\alpha \ln \biggl[ \sum\limits_{\underline{x}_\alpha}
\prod\limits_{i\in \alpha} \psi_i(x_i)
\prod\limits_{a\in \alpha} \psi_a(\underline{x}_{\partial a})
\prod\limits_{\{\mu\rightarrow \nu \ : \ \mu \in B_\alpha, \nu \in I_\alpha\}}
m_{\mu\rightarrow \nu}(\underline{x}_\nu)
\biggr] \; ,
\end{equation}
and  correction contribution
$\Delta F = -\frac{1}{\beta} \ln \bigl[ \sum_{\underline{x}}
\prod_{\alpha\in R} \omega_\alpha(\underline{x}_{\alpha})^{c_\alpha}
\bigr]$.

Let us assume that $\Delta F$
can be safely neglected in comparison with $F_0$. Then  the
free energy can be approximated as $F(\beta) \approx F_0$.
A rigorous justification
of this approximation is absent,
but it was shown in \cite{Zhou-etal-2011,Zhou-Wang-2012} that,
$\Delta F$ is the sum of correction contributions from looped region sub-graphs
at least in the cases of non-redundant
region graphs. We hope to return to this question in a future work.

The expression (\ref{eq:F0}) for $F_0$ still contains all the auxiliary
probability distribution functions. Naturally we require the free
energy $F_0$ to be stationary with respect to the chosen set of
probability functions $\{m_{\mu\rightarrow \nu}\}$.
In other words, the first derivative of $F_0$ with respect to any
$m_{\mu\rightarrow \nu}$ should be zero,
\begin{equation}
\frac{\delta F_0}{\delta m_{\mu\rightarrow \nu}} = 0 \; ,\quad\quad\quad
\forall (\mu\rightarrow \nu) \in G \; .
\end{equation}
This condition is satisfied if we require the auxiliary functions to be chosen
in such a way that, for each directed edge $\mu\rightarrow \nu$ of the region
graph $G$,
\begin{equation}
\label{eq:consist}
\sum\limits_{\underline{x}_\mu \backslash \underline{x}_\nu} \omega_\mu(
\underline{x}_\mu) \ =
\ \omega_\nu(\underline{x}_\nu) \; .
\end{equation}
Equation (\ref{eq:consist}) ensures the consistency of the two
marginal probability distributions $\omega_\mu(\underline{x}_\mu)$
and $\omega_\nu(\underline{x}_\nu)$ of each parent--child pair
$(\mu, \nu)$. Equation~(\ref{eq:omega}) and Eq.~(\ref{eq:consist})
lead to the following generalized belief-propagation equation
on each directed edge $\mu\rightarrow \nu$ of the region graph $R$:
\begin{eqnarray}
& & \hspace*{-1.0cm} \prod\limits_{\{
\alpha \rightarrow \gamma \ : \
\alpha \in B_\nu \cap I_\mu, \gamma \in I_\nu
\}}
m_{\alpha \rightarrow \gamma}(\underline{x}_\gamma )
= \nonumber \\
& & \quad\quad\quad C
\sum\limits_{\underline{x}_\mu \backslash
\underline{x}_\nu} \prod\limits_{j\in \mu \backslash \nu} \psi_j(x_j)
\prod\limits_{b\in \mu \backslash \nu} \psi_b(\underline{x}_{\partial b})
\prod\limits_{\{\eta \rightarrow \tau \ : \ \eta \in B_\mu ,
\tau \in I_\mu \backslash I_\nu\}}
 m_{\eta\rightarrow \tau}(\underline{x}_\tau) \; ,
\label{eq:gbp}
\end{eqnarray}
where $C$ is an adjustable constant to ensure that $m_{\mu\rightarrow \nu}(
\underline{x}_\nu)$ is properly normalized.
The set of GBP equations (\ref{eq:gbp})
 was first derived in \cite{Yedidia-Freeman-Weiss-2005} through
a different approach.

If the region graph $R$ is non-redundant, then for each directed
edge $\mu\rightarrow \nu$,
$B_\nu \cap I_\mu = \{\mu\}$ and $\{\alpha \rightarrow \gamma  :
 \alpha \in B_\nu \cap I_\mu, \gamma \in I_\nu\} = \{\mu\rightarrow \nu\}$.
Then equation (\ref{eq:gbp}) is simplified as
\begin{equation}
m_{\mu\rightarrow \nu}(\underline{x}_{\nu})
= C
\sum\limits_{\underline{x}_\mu \backslash
\underline{x}_\nu} \prod\limits_{j\in \mu \backslash \nu} \psi_j(x_j)
\prod\limits_{b\in \mu \backslash \nu} \psi_b(\underline{x}_{\partial b})
\prod\limits_{\{\eta \rightarrow \tau \ : \ \eta \in B_\mu ,
\tau \in I_\mu \backslash I_\nu\}}
 m_{\eta\rightarrow \tau}(\underline{x}_\tau) \; .
\label{eq:gbp1}
\end{equation}
It can be shown that equation (\ref{eq:gbp1}) is equivalent to the
region graph belief propagation (rgBP) equation of
 \cite{Zhou-etal-2011,Zhou-Wang-2012}
obtained through partition function expansion.

If the region graph $R$ is redundant, then there are multiple
directed paths between some pairs of regions. As a consequence,
the consistency conditions (\ref{eq:consist}) on some of the
directed edges must be redundant. For example, because
there are two directed paths $\alpha\rightarrow a \rightarrow i$ and
$\alpha \rightarrow d \rightarrow i$ from the square region $\alpha$ to
the vertex region $i$ in figure~\ref{fig:2Dgroup:b},
then if the consistency conditions on the
edges $\alpha \rightarrow a$, $\alpha \rightarrow d$ and $a \rightarrow i$
are all satisfied, the consistency condition on the edge $d\rightarrow i$
is satisfied automatically. For a redundant region graph, some of the
consistency conditions (\ref{eq:consist}) can therefore be dropped
without affecting the final theoretical results. Consequently, some of
the GBP equations (\ref{eq:gbp}) do not need to be considered.

\section{Simplified GBP (SGBP) on a redundant
region graph}
\label{sec:sgbp}

In this paper we are interested in the case of the region graph $R$ being
redundant. For example, figure~\ref{fig:2Drg:b} shows
a redundant region graph for the 2D Edwards-Anderson model on a square lattice
with period boundary conditions. In this region graph, the region
subgraph $R_i$ induced by each vertex $i$ is not a tree.
It contains nine regions and twelve directed edges, and loops exist in
$R_i$ at the level of regions.

According to the definition (\ref{eq:zalpha}), each boundary edge
 $(\mu\rightarrow \nu)$
of region $\alpha$ contributes a product term
 $m_{\mu\rightarrow \nu}(\underline{x}_\nu)$
to the Boltzmann factor $z_\alpha$. As an example, consider the rod region
$a$ of figure~\ref{fig:2Dgroup:b}. Its Boltzmann factor $z_a$ is
\begin{eqnarray}
z_a &= & \sum_{x_i, x_j} \psi_i(x_i) \psi_j(x_j) \psi_a(x_i, x_j)
m_{\alpha \rightarrow a}(x_i,x_j) m_{\beta\rightarrow a}(x_i,x_j)
m_{h\rightarrow i}(x_i) m_{f\rightarrow j}(x_j) \nonumber \\
& & \quad\quad\quad\quad\quad\quad \times
m_{d\rightarrow i}(x_i) m_{g\rightarrow i}(x_i)
 m_{b\rightarrow j}(x_j) m_{e\rightarrow j}(x_j)
\; .
\label{eq:zaexample}
\end{eqnarray}
Notice that the square region $\alpha$ and its two child rod regions $b$
and $d$ all
send a message to region $a$ or its descendants, and these three
messages are assumed to be mutually independent in the
probability product form of equation (\ref{eq:zaexample}). Similarly, the
 square
region $\beta$ and its two child rod regions $e$ and $g$ contribute a product
term of three probability messages to $z_a$.
However, because the rod regions $b$ and $d$ are children of region $\alpha$,
 the probability message $m_{\alpha \rightarrow a}(x_i, x_j)$ already
 contains the effects
of region $b$ and $d$ to region $a$. In other words, the probability
distributions $m_{d\rightarrow i}(x_i)$ and
$m_{b\rightarrow j}(x_j)$  very likely are strongly related to the probability
distribution $m_{\alpha\rightarrow a}(x_i, x_j)$. Similarly, we expect the
probability distributions $m_{g\rightarrow i}(x_i)$ and $m_{e\rightarrow j}(x_j)$
to be strongly related to the probability
distribution $m_{\beta \rightarrow a}(x_i, x_j)$.

All such possibly strong dependence effects among the input probability
distributions are completely neglected in (\ref{eq:zaexample}) and in
the more general expression (\ref{eq:zalpha}).
In this sense the redundance in the region graph structure
causes redundance in the
Boltzmann factor expressions (\ref{eq:zalpha}), which in turn contributes at least
part of the redundance in the set of GBP equations (\ref{eq:gbp}) and increases
the complexity of numerical computations. The issue of removing the
GBP redundance has been discussed by several
recent studies
\cite{Zhou-Wang-2012,Rizzo-etal-2010,LageCastellanos-etal-2011,Dominguez-etal-2011,LageCastellanos-etal-2012}.

We now propose a new way of removing redundance in the GBP equations.
The basic idea is very simple: for a region $\nu$ in the boundary set $B_\alpha$
of region $\alpha$, the input probability messages
from $\nu$ to the set $I_\alpha$ should
be removed as many as possible if an ancestor of $\nu$ also belongs to the
boundary set $B_\alpha$. Let us first discuss the ideal situation,
namely the following equality is valid on each directed edge
$\mu \rightarrow \nu$ of the region graph $R$:
\begin{equation}
\label{eq:e-identity}
\sum\limits_{\{\alpha \ : \ \mu \in B_\alpha, \nu \in I_\alpha,
A_\mu \cap B_\alpha = \emptyset \}}  c_\alpha
= 0 \; ,
\end{equation}
where $A_\mu$ is the ancestor set of region $\mu$ as defined in section~\ref{subsec:rg}.
The summation on the left-hand side of the above expression is over all the
regions $\alpha$ satisfying the following conditions:
(1) the boundary set $B_\alpha$ contains region $\mu$ but
not any of the ancestors of $\mu$ ($A_\mu \cap B_\alpha = \emptyset$);
and (2) the interior set $I_\alpha$ contains region $\nu$. As a simple example,
let us mention that
identity (\ref{eq:e-identity}) is satisfied on each directed edge of the
redundant region graph of figure~\ref{fig:2Drg:b}. This particular region graph is
therefore redundant and `ideal'.

If (\ref{eq:e-identity}) holds on each directed edge $\mu\rightarrow \nu$,
then the partition function expression (\ref{eq:Z-exp-2}) can be simplified as
\begin{equation}
Z(\beta)  =
\sum\limits_{\underline{x}}
\prod\limits_{\alpha\in R}
\biggl[
\prod\limits_{i\in \alpha} \psi_i(x_i)
\prod\limits_{a\in \alpha} \psi_a(\underline{x}_{\partial a})
\prod\limits_{\{\mu\rightarrow \nu \
: \
\mu\in B_\alpha, \nu\in I_\alpha, A_\mu \cap B_\alpha = \emptyset
\}} m_{\mu\rightarrow \nu}(\underline{x}_\nu)
\biggr]^{c_\alpha}  \; .
\label{eq:Z-exp-simp}
\end{equation}
We can then define a simplified Boltzmann factor for region $\alpha$ as
\begin{equation}
\tilde{z}_\alpha  \equiv \sum\limits_{\underline{x}_\alpha}
\prod\limits_{i\in \alpha} \psi_i(x_i)
\prod\limits_{a\in \alpha} \psi_a(\underline{x}_{\partial a})
\prod\limits_{\{\mu\rightarrow \nu \ : \
 \mu \in B_\alpha, \nu \in I_\alpha, A_\mu \cap B_\alpha = \emptyset \}}
m_{\mu\rightarrow \nu}(\underline{x}_\nu) \; .
\end{equation}
A simplified Boltzmann weight for region $\alpha$ is defined correspondingly:
\begin{equation}
\tilde{\omega}_\alpha(\underline{x}_\alpha)
  \equiv \frac{1}{\tilde{z}_\alpha} \prod\limits_{i\in \alpha} \psi_i(x_i)
\prod\limits_{a\in \alpha} \psi_a(\underline{x}_{\partial a})
\prod\limits_{\{\mu\rightarrow \nu \ : \ \mu \in B_\alpha, \nu \in I_\alpha,
A_\mu \cap B_\alpha = \emptyset \}}
m_{\mu\rightarrow \nu}(\underline{x}_\nu) \; .
\end{equation}
For the ideal redundant region graph of figure~\ref{fig:2Drg:b},
the Boltzmann factor $\tilde{z}_a$ of the rod region
$a$ is readily written down as (see figure~\ref{fig:2Dgroup:b})
\begin{equation}
\tilde{z}_a =
 \sum\limits_{x_i, x_j} \psi_i(x_i) \psi_j(x_j) \psi_a(x_i, x_j)
m_{\alpha \rightarrow a}(x_i,x_j) m_{\beta\rightarrow a}(x_i,x_j)
m_{h\rightarrow i}(x_i) m_{f\rightarrow j}(x_j) \; ,
\label{eq:zaexample2}
\end{equation}
which is much simpler than the expression (\ref{eq:zaexample}).

The partition function is then expressed as
\begin{equation}
Z(\beta) = \tilde{Z}_0 \times \sum\limits_{\underline{x}} \prod\limits_{\alpha}
\tilde{\omega}_\alpha (\underline{x}_\alpha)^{c_\alpha} \; ,
\end{equation}
where $\tilde{Z}_0 \equiv \prod\limits_{\alpha} \tilde{z}_\alpha^{c_\alpha}$.
A new approximate free energy $\tilde{F}_0$ is defined as
\begin{equation}
\label{eq:tildeF0}
\tilde{F}_0 \equiv -
\frac{1}{\beta} \ln \tilde{Z}_0
= -\frac{1}{\beta} \sum\limits_{\alpha \in R}
c_\alpha \ln \biggl[ \sum\limits_{\underline{x}_\alpha}
\prod\limits_{i\in \alpha} \psi_i(x_i)
\prod\limits_{a\in \alpha} \psi_a(\underline{x}_{\partial a})
\prod\limits_{\{\mu\rightarrow \nu\ : \ \mu \in B_\alpha, \nu \in I_\alpha,
A_\mu \cap B_\alpha = \emptyset \}}
m_{\mu\rightarrow \nu}(\underline{x}_\nu)
\biggr] \; .
\end{equation}
Requiring $\tilde{F}_0$ to be stationary with respect to the set
of auxiliary probability distributions $\{m_{\mu\rightarrow \nu}(\underline{x}_\nu)\}$,
the following set of simplified consistency conditions are obtained:
\begin{equation}
\label{eq:tildeconsist}
\sum\limits_{\underline{x}_\mu \backslash \underline{x}_\nu}
\tilde{\omega}_\mu(
\underline{x}_\mu) \ =
\ \tilde{\omega}_\nu(\underline{x}_\nu) \; , \quad\quad\quad
\forall (\mu\rightarrow \nu) \in R \; .
\end{equation}
This consistency condition leads to the following simplified generalized
belief propagation (SGBP) equation on each directed edge $\mu\rightarrow \nu$:
\begin{eqnarray}
& & \hspace*{-1.0cm} \prod\limits_{\{\alpha \rightarrow \gamma \ : \
\alpha \in B_\nu \cap I_\mu, \gamma \in I_\nu, A_\alpha \cap B_\nu = \emptyset \}}
m_{\alpha \rightarrow \gamma}(\underline{x}_\gamma ) = \nonumber \\
& & \quad\quad\quad\quad
C \sum\limits_{\underline{x}_\mu \backslash \underline{x}_\nu }
 \prod\limits_{j\in \mu \backslash \nu} \psi_j(x_j)
\prod\limits_{b\in \mu \backslash \nu} \psi_b(\underline{x}_{\partial b})
\prod\limits_{\{\eta \rightarrow \tau \ : \
\eta \in B_\mu,
\tau \in I_\mu \backslash I_\nu, A_\alpha \cap B_\mu = \emptyset\}}
m_{\eta\rightarrow \tau}(\underline{x}_\tau) \; ,
\label{eq:sgbp}
\end{eqnarray}
where $C$ is again an adjustable normalisation constant.

For a given spin glass model (\ref{eq:totalenergy}), it may be a
relatively easy task to construct a redundant region graph that has the property
(\ref{eq:e-identity}) on each of its directed edges. The set of
SGBP equations (\ref{eq:sgbp}) can then be iterated on such an `ideal' redundant
region graph $R$, and the approximate free energy $\tilde{F}_0$ can be
computed accordingly.

To be complete, let us also briefly discuss the `non-ideal' case in which
the identity (\ref{eq:e-identity}) holds on some but not all of
the directed edges. A simple solution would be to divide the directed edges
into two classes (say $C_I$ and $C_N$), one ($C_I$)
 with the property (\ref{eq:e-identity}) and the other
one ($C_N$) with only the weaker property (\ref{eq:c-identity}). For
an edge $(\mu\rightarrow \nu)\in C_I$, the probability message
$m_{\mu\rightarrow \nu}(\underline{x}_\nu)$ is received by all the
regions $\alpha$ in the set $\{\alpha: \mu \in B_\alpha, \nu \in I_\alpha,
A_\mu \cap B_\alpha = \emptyset\}$. While for an edge $(\eta \rightarrow \tau) \in
C_N$, we require the probability message
$m_{\eta\rightarrow \tau}(\underline{x}_\tau)$ to be received by all
the regions in a properly constructed non-empty subset of the
set $\{\gamma: \eta \in B_\gamma, \tau \in I_\gamma\}$ (this
constructed subset has the property that the
sum of counting numbers of its regions is equal to zero).
Following the theoretical approach of this section we can then obtain a
modified set of SGBP equations for the non-ideal region graph $R$.

As there are multiple directed paths between some pairs of regions in
the redundant region graph $R$, the simplified
consistency conditions (\ref{eq:tildeconsist}) on some of the
directed edges must be redundant (see the discussion at the last
paragraph of section~\ref{sec:pfrg}).
Therefore some of these simplified consistency conditions and
the corresponding SGBP equations do not need to be considered in the
actual numerical iteration process.

\section{Applications to the Ising and the
Edwards-Anderson model}
\label{sec:EA}

It is helpful to complement the theoretical discussions of the preceding
section with some applications.
We now apply the SGBP equations to the ferromagnetic Ising model and the
Edwards-Anderson spin glass model (\ref{eq:EAh})
on a square lattice or a cubic lattice. Periodic boundary condition
is assumed on each dimension of the lattice systems.
For the Ising model, each coupling constant $J_{i j} \equiv J$, while
for the EA model $J_{i j}$ is assigned a value $+J$ or $-J$ independently
and uniformly at random and then is fixed in time
(we shall set $J=1$ in the following discussions).

For the ideal redundant region graph (later referred to as
$R_{2D}^{n=2}$) of figure~\ref{fig:2Drg:b},
the simplified
Boltzmann weights of a square region $\alpha$, a rod region $a$ and a
vertex region $i$ can be easily written down as (see figure~\ref{fig:2Dgroup})
\numparts
\begin{eqnarray}
\fl
\tilde{\omega}_{\alpha}(x_i, x_j, x_k, x_l)
&\propto & e^{\beta (h_{i}^0 x_{i}
+h_{j}^0 x_{j} + h_{k}^0 x_{k} + h_{l}^0 x_{l} +
 J_{i j} x_{i} x_{j}
+ J_{j k} x_{j} x_{k} + J_{k l} x_{k} x_{l}
 + J_{l i} x_{l} x_{i})} \nonumber \\
   & & \quad\quad\quad \times
m_{\beta \rightarrow a}(x_{i}, x_{j})
m_{\nu \rightarrow b}(x_{j}, x_{k})
m_{\mu \rightarrow c}(x_{k}, x_{l})
m_{\gamma \rightarrow d}(x_{l}, x_{i}) \; ,
 \\
\tilde{\omega}_{a}(x_i, x_j) & \propto &
e^{\beta ( h_i^0 x_i + h_j^0 x_j + J_{i j} x_i x_j) }
m_{\alpha \rightarrow a}(x_i, x_j) m_{\beta \rightarrow a}(x_i, x_j)
m_{h \rightarrow i}(x_i) m_{f \rightarrow j}(x_j) \; ,
\\
\tilde{\omega}_i(x_i) & \propto &
e^{\beta h_i^0 x_i}
m_{a\rightarrow i}(x_i) m_{h\rightarrow i}(x_i)
m_{d\rightarrow i}(x_i) m_{g\rightarrow i}(x_i) \; .
\end{eqnarray}
\endnumparts
The SGBP equations on the directed edges $\alpha\rightarrow a$ and
$a\rightarrow i$ are, respectively
\numparts
\begin{eqnarray}
\label{eq:sgbp1}
 m_{\alpha \rightarrow a}(x_i, x_j) m_{h\rightarrow i}(x_i)
m_{f\rightarrow j}(x_j)  \propto
\sum\limits_{x_k, x_l} e^{\beta (h_k^0 x_k + h_l^0 x_l + J_{j k} x_j x_k
+J_{k l} x_k x_l + J_{l i} x_l x_i)} \quad\quad\quad\quad\quad \nonumber \\
\quad\quad\quad\quad\quad \times m_{\nu \rightarrow b}(x_j, x_k)
m_{\mu \rightarrow c}(x_k, x_l)
m_{\gamma \rightarrow d}(x_l, x_i) \; ,
\\
\label{eq:sgbp2}
\hspace*{-0.8cm}  m_{a\rightarrow i}(x_i) m_{d\rightarrow i}(x_i)
m_{g\rightarrow i}(x_i) \propto
\sum\limits_{x_j} e^{\beta (h_j^0 x_j +  J_{i j} x_i x_j)}
 m_{\alpha \rightarrow a}(x_i, x_j) m_{\beta\rightarrow a}(x_i,x_j)
m_{f\rightarrow j}(x_j)\; .
\end{eqnarray}
\endnumparts
\begin{figure}
\includegraphics[width=0.31\textwidth]{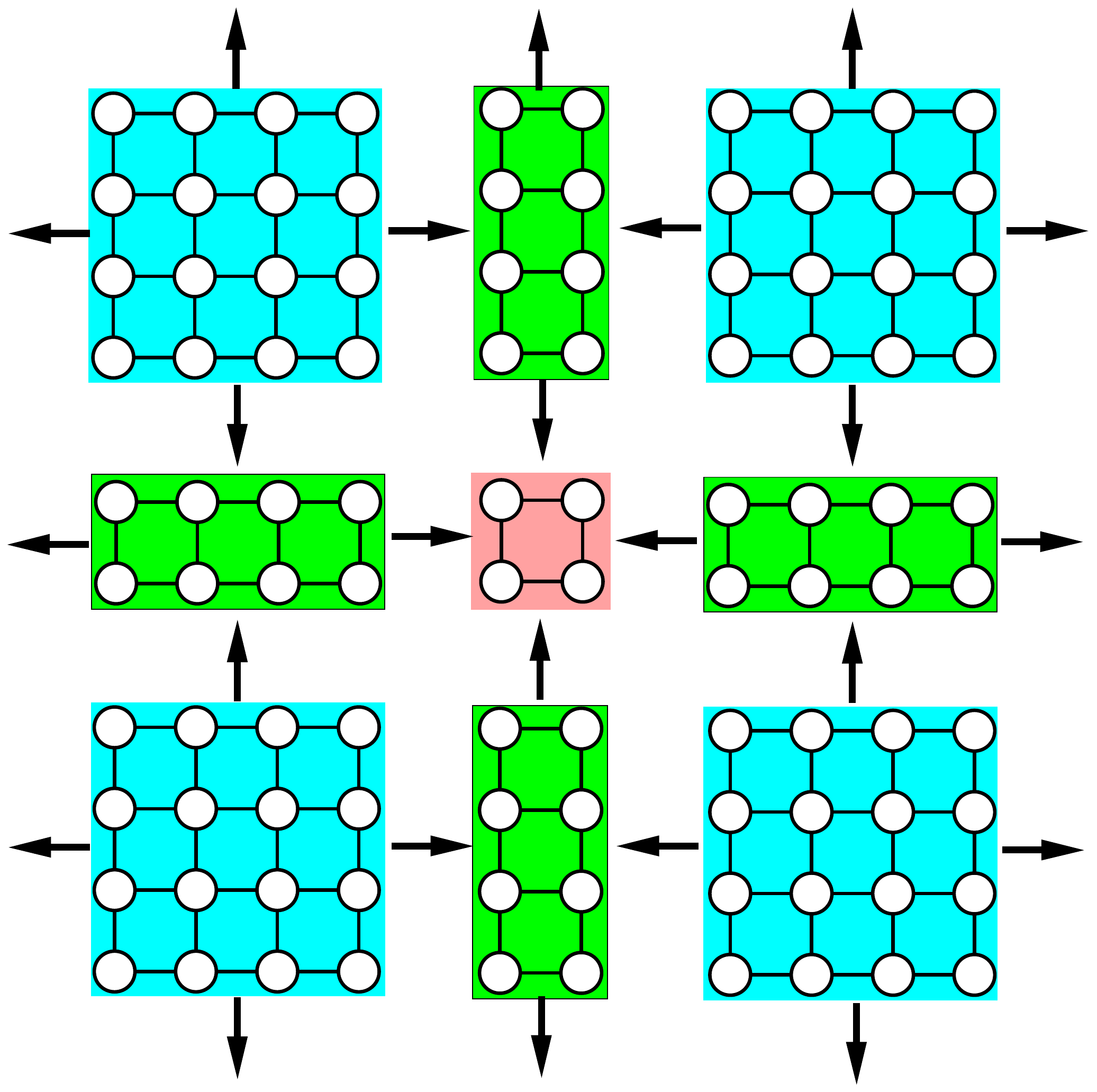}\hspace{2pc}%
\begin{minipage}[b]{24pc}
\caption{\label{fig:rg_gbp4}
The local structure of another redundant region graph for the 2D model of
figure~\ref{fig:2Drg:a}. Each square region (cyan color)
contains $4 \times 4$ vertices and is parental to four rod regions,
its counting number is $c=1$; each rod region (green color) contains
$2 \times 4$ vertices and is parental to two plaquette regions,
its counting number is
$c=-1$; each plaquette region (pink color)
contains $2\times 2$ vertices, its counting number is $c=1$. For
simplicity we do not show the small squares of
figure~\ref{fig:2Drg:a} here but use a small edge between two
vertices to indicate a coupling interaction.
}
\end{minipage}
\end{figure}

To include  more local correlations in the SGBP equations, we also
consider another ideal redundant region graph (referred to as
$R_{2D}^{n=4}$, see
figure~\ref{fig:rg_gbp4}) for the 2D system (\ref{eq:EAh}).
This region graph has the same topology as $R_{2D}^{n=2}$.
The only difference is that each vertex region now
becomes a plaquette region of $2\times 2$ vertices, and each square
region now contains $4\times 4$ vertices. The SGBP equations for this
region graph are similar to equations (\ref{eq:sgbp1})
and (\ref{eq:sgbp2}).

For the cubic lattice systems we consider a simple 3D
extension of the region graph of figure~\ref{fig:2Drg:b}.
This extended region graph (later referred to as $R_{3D}^{n=2}$)
has four types of regions: cube regions, surface regions,
rod regions, and vertex regions.
Each $2\times 2\times 2$ cubic cell of the 3D lattice forms a cube region of
the region graph, it is parental to six surface regions,
and its counting number is $c=1$;
each $2\times 2$ plaquette of the lattice
forms a surface region, it is parental to four rod regions and
has counting number $c=-1$; each pair of nearest neighboring
vertices of the lattice forms  a rod region (it is
parental to two vertex regions, and
its counting number is $c=1$); and each vertex region
contains only one single vertex, with counting number $c=-1$. The
SGBP equations for such a 3D region graph are much more simplified in
comparison with the original GBP equations.

We are mainly interested in the spontaneous emergence of collective
behaviors, therefore we set the external field on each vertex
to be zero ($h_i^0 =0$ for all the vertices $i$).

\subsection{The 2D Ising model}

\begin{figure}
\centering
\subfigure[]{
\label{fig:2DIsing:a}
\includegraphics[width=0.45\textwidth]{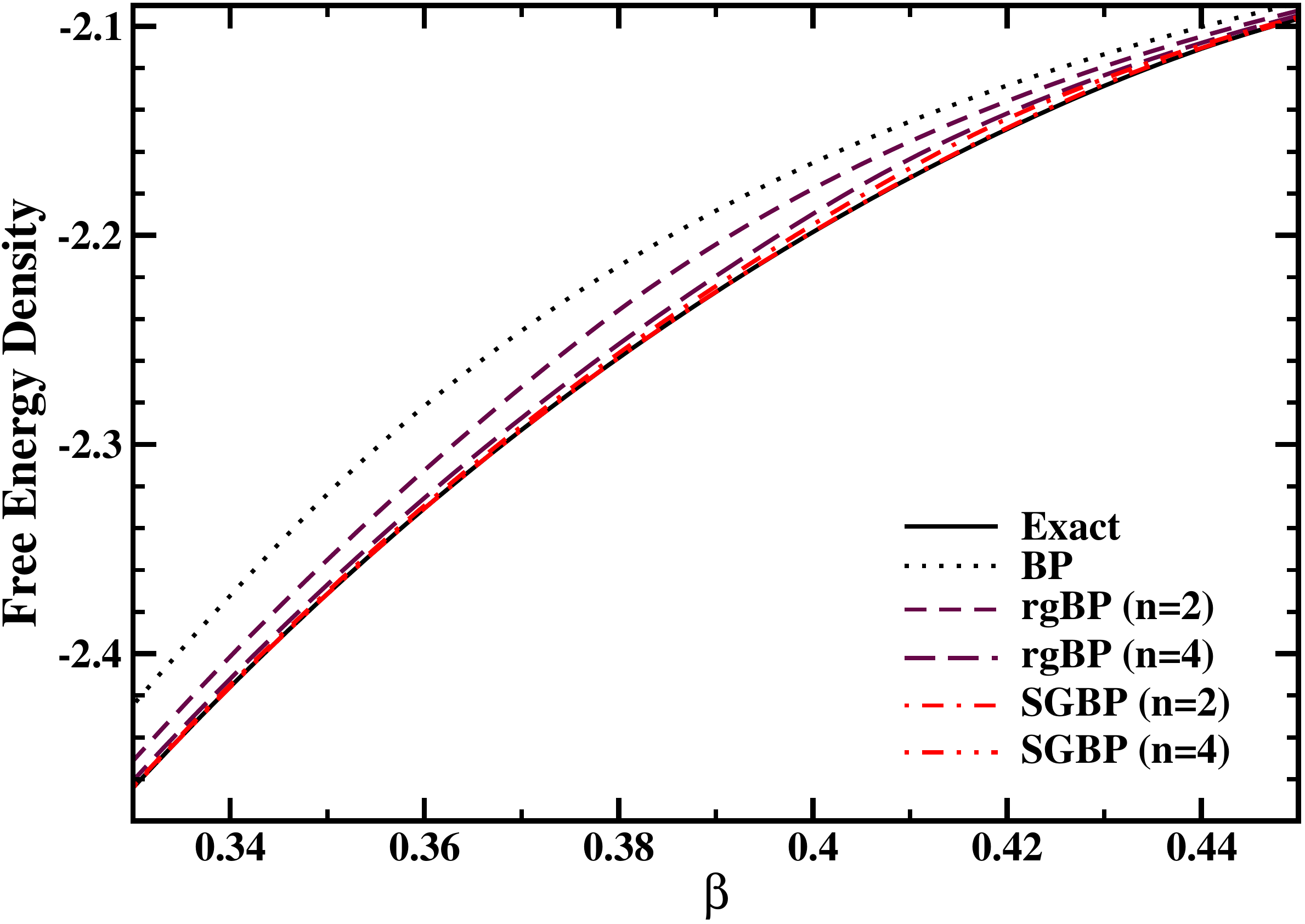}}
\hspace{2pc}
\subfigure[]{
\label{fig:2DIsing:b}
\includegraphics[width=0.45\textwidth]{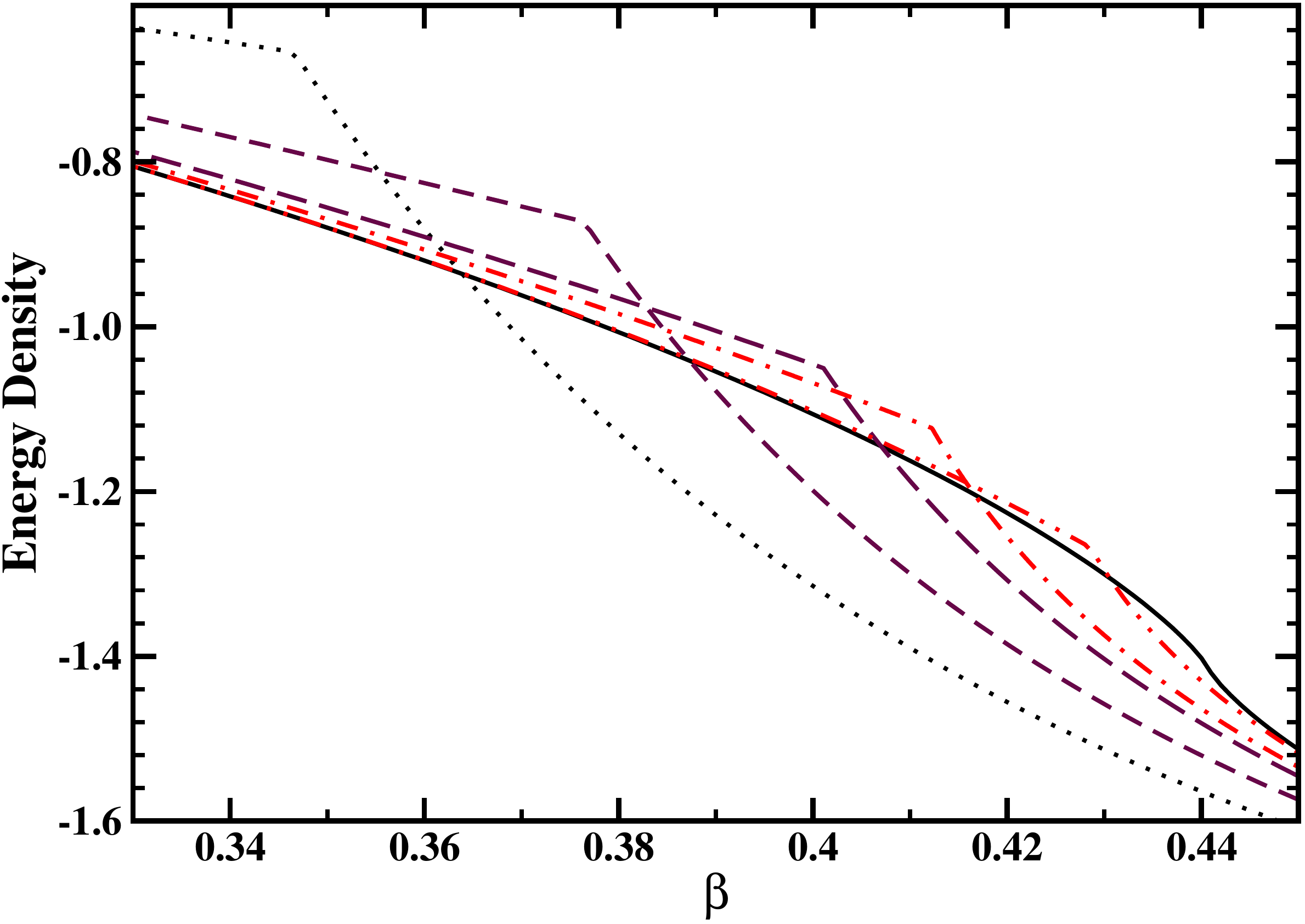}}
\subfigure[]{
\label{fig:2DIsing:c}
\includegraphics[width=0.45\textwidth]{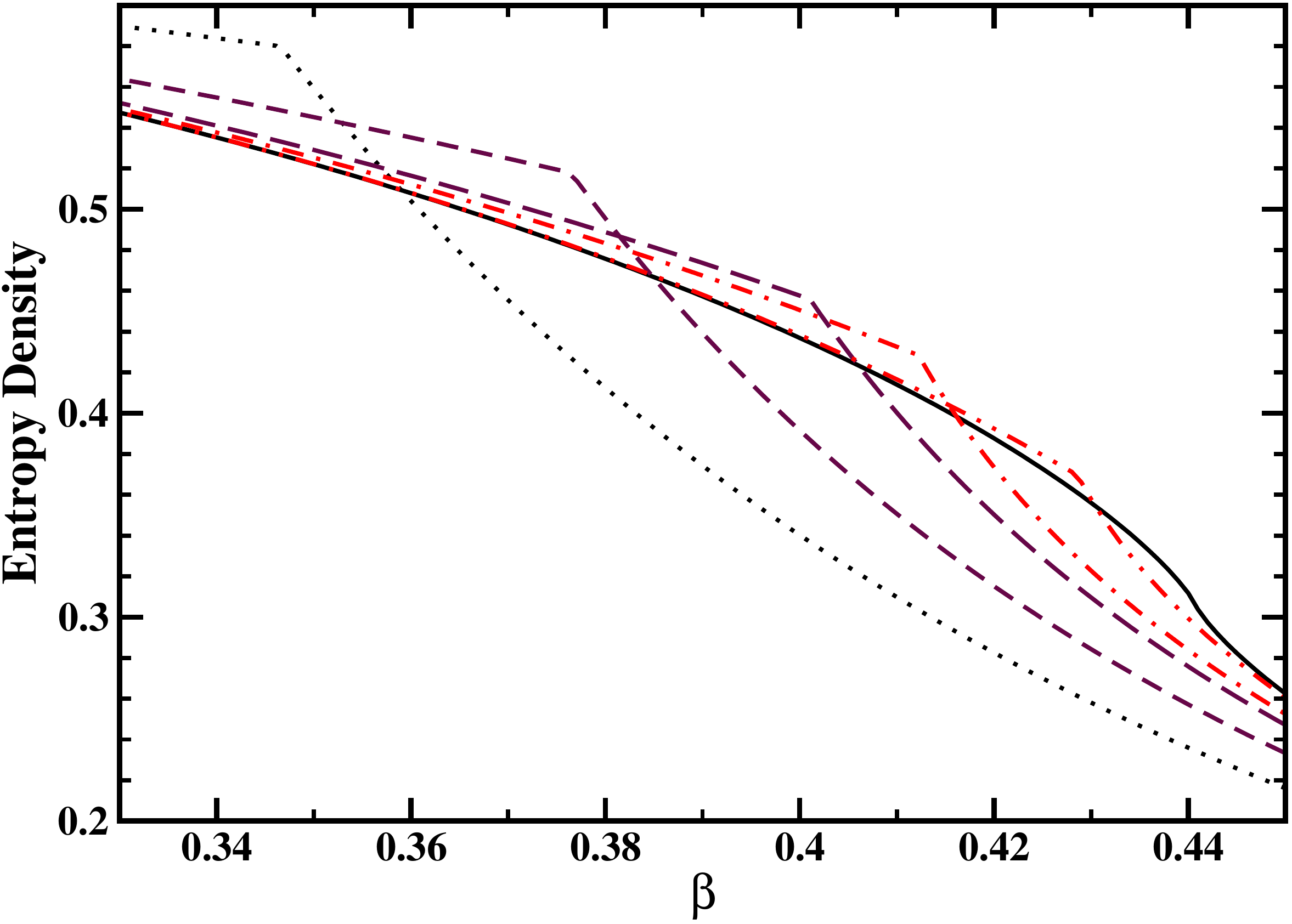}}
\hspace{2pc}
\subfigure[]{
\label{fig:2DIsing:d}
\includegraphics[width=0.45\textwidth]{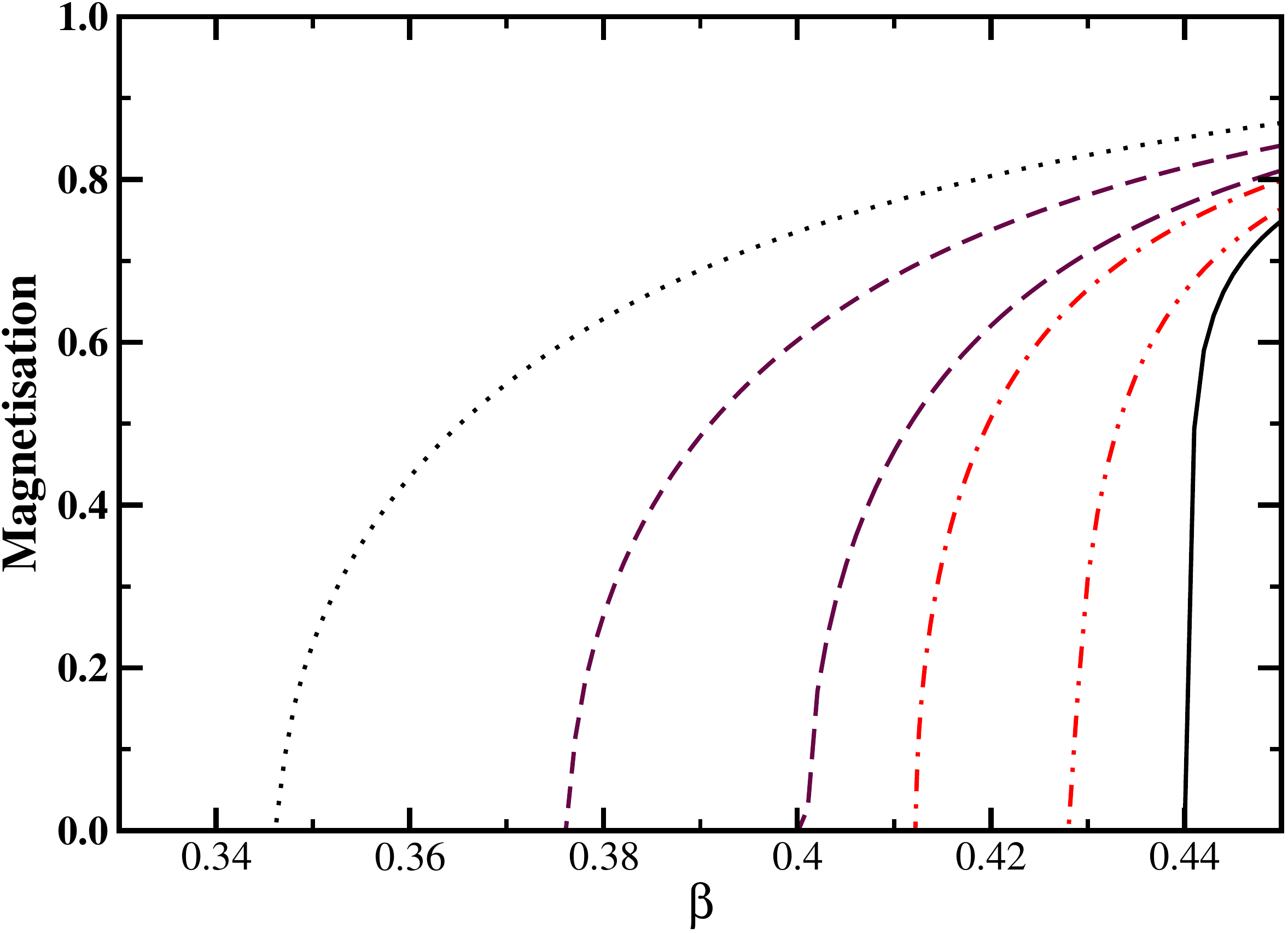}}
\caption{Results on the 2D ferromagnetic Ising model obtained by the SGBP
equations, the rgBP equations of \cite{Zhou-Wang-2012}, and
the conventional BP equation: (a) free energy density; (b) mean energy
density; (c) entropy density; (d) magnetisation (mean spin value of a
single vertex). The solid lines denote the exact results of
\cite{Onsager-1944}. The parameter $n$ of the SGBP equations and the rgBP
equations is the number of vertices on each side of the maximal square
region of the region graph. For the SGBP curves, $n=2$ and $n=4$
correspond to
the region graph of figure~\ref{fig:2Drg:b} and
figure~\ref{fig:rg_gbp4}, respectively. The region graphs at $n=2$
and $n=4$ for the
rgBP equations are described in \cite{Zhou-Wang-2012}.
}
 \label{fig:2DIsing}
\end{figure}
First we consider the simplest case, namely the ferromagnetic Ising
model on an infinite square lattice with periodic boundary conditions.
As the system has no disorder the SGBP equations can be easily solved
at all temperatures.
In figure~\ref{fig:2DIsing} we compare the results obtained by the
SGBP equations
with the exact results of Onsager \cite{Onsager-1944} and with the results
obtained through the region graph belief propagation (rgBP) equations
(see \cite{Zhou-Wang-2012,Zhou-etal-2011} for details).

For the SGBP equations on a given region graph, we perform linear
stability analysis to determine precisely
the instability point of the paramagnetic solution,
similar to that performed in \cite{Zhou-Wang-2012}.
The predicted transition inverse temperatures
by the SGBP equations using
the region graph $R_{2D}^{n=2}$ and $R_{2D}^{n=4}$
are, respectively, $\beta_c \approx 0.4126$ and $\beta_c \approx 0.429$.
These values are rather close to the
exactly known
 critical inverse temperature value of $\beta_c^{(exact)} \approx 0.4407$
\cite{Kramers-Wannier-1941a},
they are much improved over the predicted
value of $\beta_c \approx 0.3466$ by the conventional BP method and
the predicted values of $\beta_c\approx 0.3765$ and $\beta_c \approx 0.401$
by the rgBP equations on two non-redundant region graphs
 \cite{Zhou-Wang-2012}.

The SGBP equations also give better predictions on the free energy
density, the mean energy density, the entropy density, and the
mean magnetisation of the system, see figure~\ref{fig:2DIsing}.
Using the region graph $R_{2D}^{n=4}$,
the predicted values of the SGBP
equations on the free energy density, energy density, and
entropy density become very close to the exactly known results.
The prediction power of the SGBP equations will be further improved if
we consider more local correlations by enlarging
 each region of figure~\ref{fig:rg_gbp4} while keeping unchanged the
topology of this region graph.

\subsection{The 2D Edwards-Anderson model}

Because of the existence of multiple directed paths between some pairs of
regions in the redundant region graph $R_{2D}^{n=2}$ or $R_{2D}^{n=4}$,
the SGBP equations on some of the directed edges are redundant and can be
dropped from the numerical iteration process (see discussion in the last
paragraph of section~\ref{sec:sgbp}).
If instead the SGBP equations on all the
directed edges of the region graph are used in the iteration process,
the paramagnetic solution of the SGBP equations will be unstable even
at $\beta=0$ ($T=\infty$).

In our numerical implementations for the 2D Edwards-Anderson model
we still keep the SGBP equations on
all the directed edges but add a damping effect to the SGBP equations to
facilitate the convergence.
We notice that the paramagnetic solution of the SGBP
equations on the region graphs $R_{2D}^{n=2}$ and $R_{2D}^{n=4}$ fail to
be stable at low temperatures even for optimally adjusted  magnitude of
the damping effect.
However, the $\pm J$ EA model on
2D square lattice is believed to have no finite-temperature spin glass
phase
\cite{Thomas-Huse-Middleton-2011}.
Therefore only the paramagnetic solution of the SGBP equations are
qualitatively correct, even if it is unstable at low temperatures.

\begin{figure}
\includegraphics[width=0.5\textwidth]{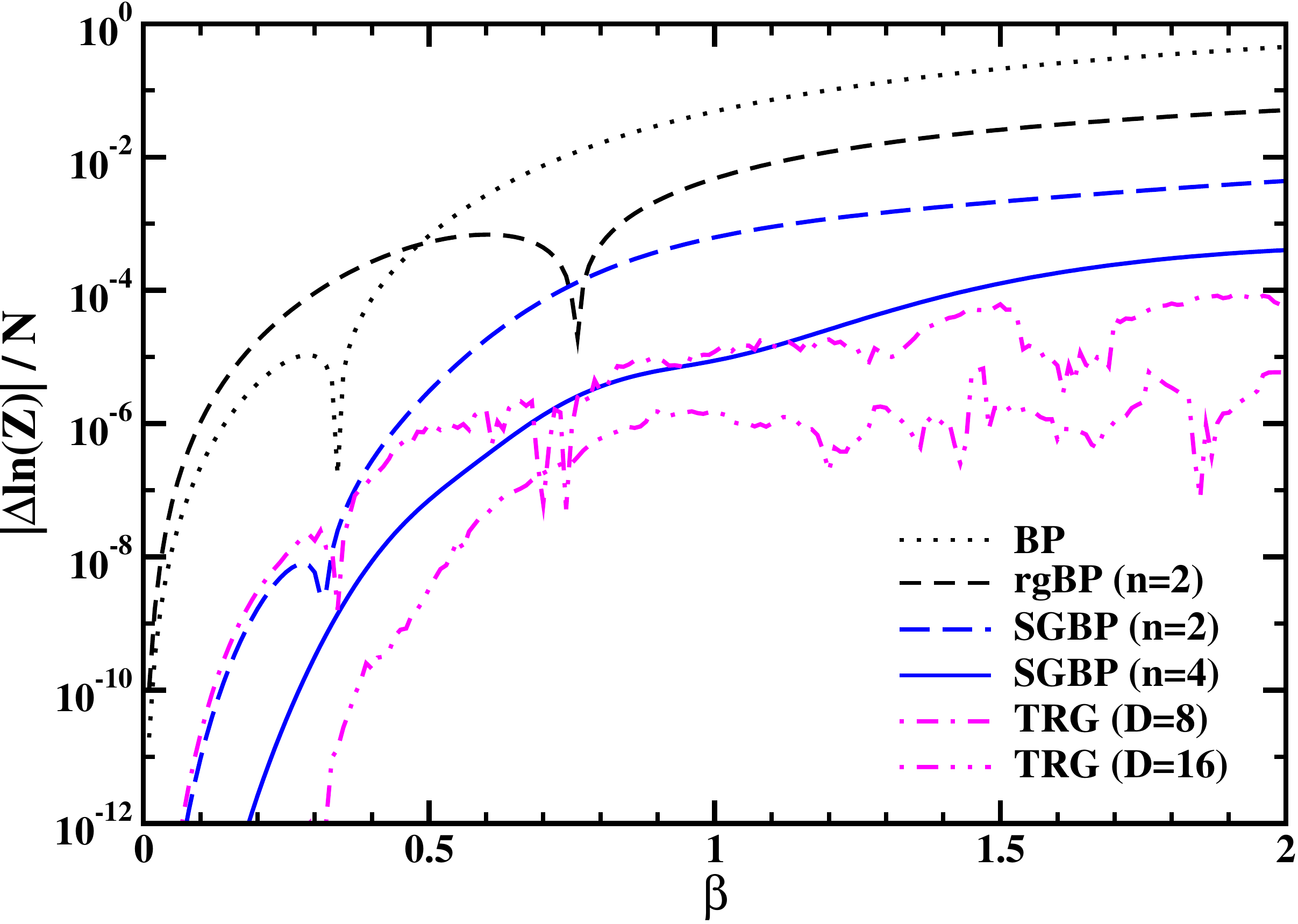}\hspace{2pc}%
\begin{minipage}[b]{16.5pc}
\caption{\label{fig:2DEAerror}
Difference $\Delta \ln Z$ between the exactly calculated value of
$\ln Z$ and the value of $\ln Z$ obtained by BP, rgBP with $n=2$,
SGBP with $n=2$ and
$n=4$, and tensor renormalisation group (TRG) with cutoff parameter $D=8$ and
$D=16$. All the numerical calculations are performed on a single instance
of the 2D EA model with side length
$L=64$ and periodic boundary conditions (the total number of vertices is
$N=64^2$).
}
\end{minipage}
\end{figure}

Since the free energy of a given instance of the EA model on a
periodic square lattice with side length $L$ can be computed exactly
in polynomial
time \cite{Barahona-1982}, we can compare the free energy
$\tilde{F}_0$ of the paramagnetic
solution of the SGBP equations with the exact value,
see figure~\ref{fig:2DEAerror}. The SGBP results on the region graph
$R_{2D}^{n=4}$ are comparable to the results obtained by
the tensor renormalisation group (TRG) method
\cite{Levin-Nave-2007} with cutoff parameter $D=8$ but are
worse than the TRG results with cutoff parameter $D=16$. We expect that
the performance of
the SGBP equations will be further improved by enlarging the side length
$n$ of the square regions.
An advantage of the SGBP message passing approach is that
it can simultaneously give the marginal
probability distributions $\tilde{\omega}_\mu
(\underline{x}_\mu)$ for all the regions $\mu$.

\subsection{The 3D Ising model and Edwards-Anderson
model}

The structure of a simple redundant region graph
$R_{3D}^{n=2}$ for the 3D Ising model and
EA model is described at the beginning of
section~\ref{sec:EA}. This region graph is formed by
cube regions, surface regions,
rod regions and vertex regions. Each  cube region contains
$n \times n \times n$ vertices with $n=2$.
Linear stability analysis of the SGBP equations
predicts that the ferromagnetic transition occurs at
the critical inverse temperature $\beta_c \approx 0.2183$, which
is very close to the value of $\beta_c \approx 0.2217$ obtained by
Monte Carlo simulations \cite{Talapov-Blote-1996}
and the higher-order tensor renormalisation group (TRG) method
\cite{Xie-etal-2012}. As a comparison, the critical inverse
temperature predicted by the
BP approximation is $\beta_c \approx 0.2027$.

\begin{figure}
\begin{minipage}{18pc}
\includegraphics[width=18pc]{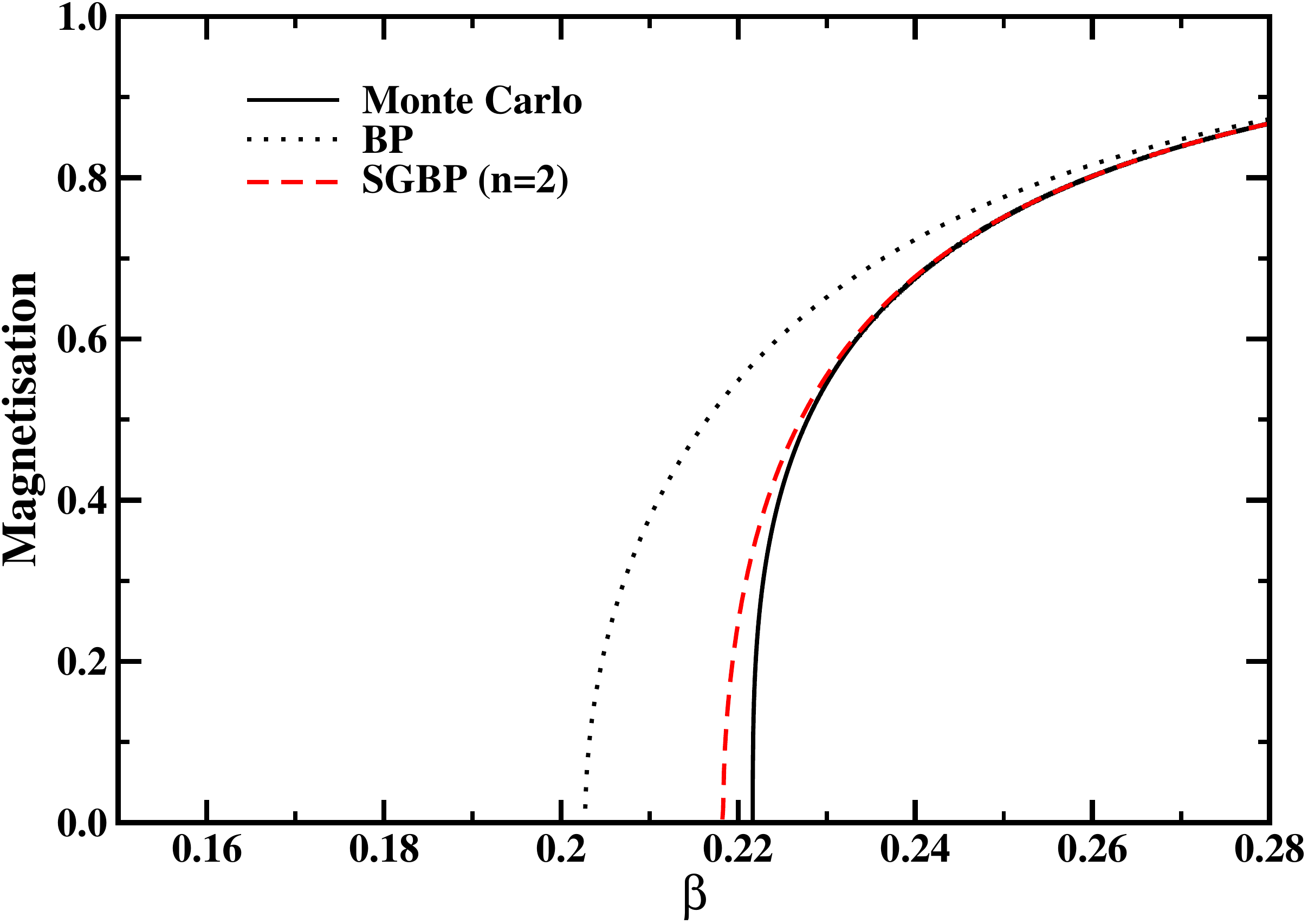}
\caption{
\label{fig:3DIsing}
Mean magnetisation of the 3D Ising model as obtained by
the SGBP equations (using region graph $R_{3D}^{n=2}$), the
BP equations, and Monte Carlo
simulations \cite{Talapov-Blote-1996}.}
\end{minipage}\hspace{2pc}%
\begin{minipage}{18pc}
\includegraphics[width=18pc]{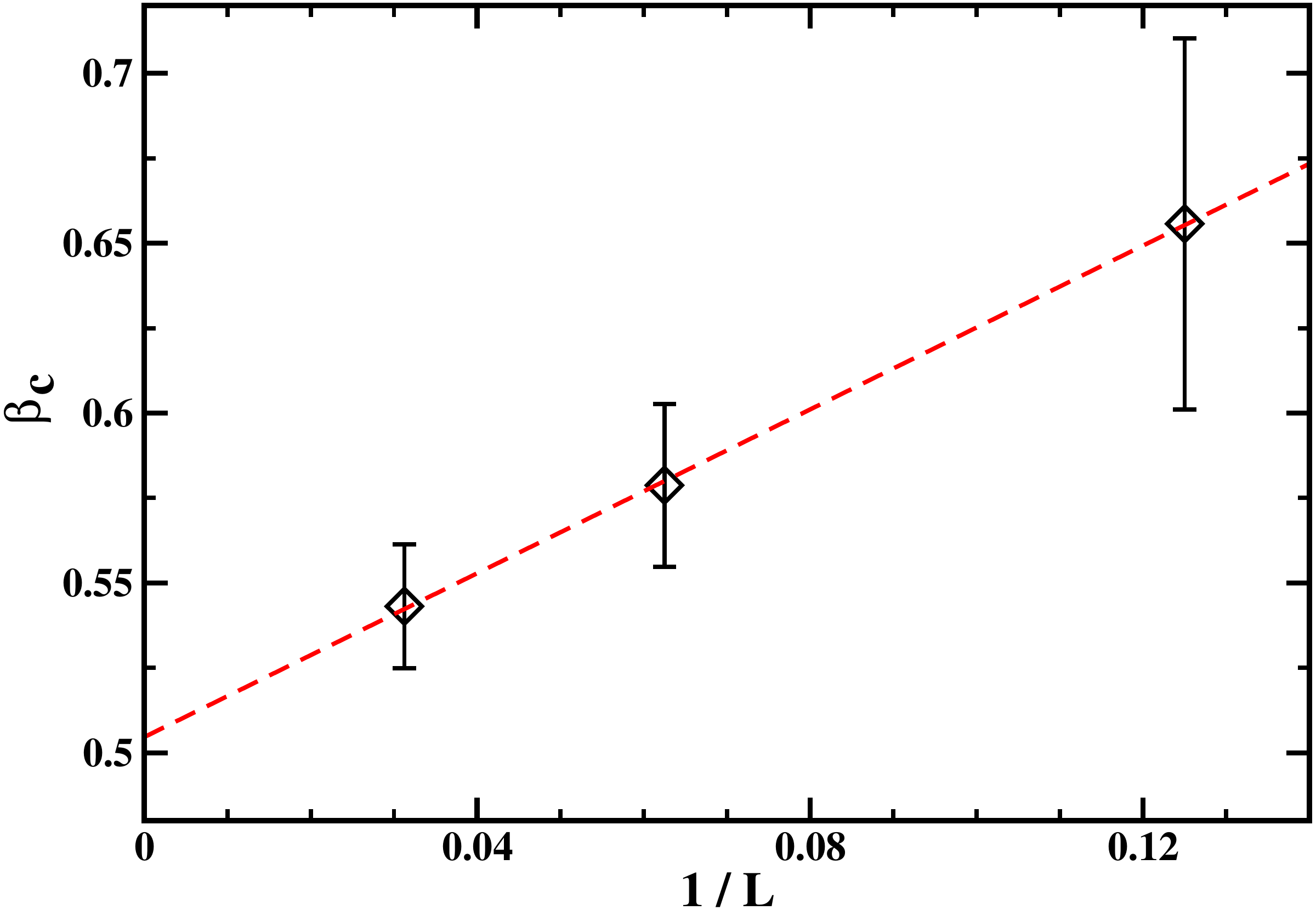}
\caption{\label{fig:3DEA} The inverse temperature $\beta_c$
for the instability of the paramagnetic solution of the SGBP equations on
the 3D EA model with lattice side length $L$. The dashed line is
the fitting curve of $\beta_c =
0.505 + 1.20/L$.}
\end{minipage}
\end{figure}

The magnetisation values predicted  by the
BP and SGBP equations are compared with the Monte Carlo simulation
results of \cite{Talapov-Blote-1996} in figure~\ref{fig:3DIsing}.
It is evident that the performance of SGBP  improves
considerably over that of BP.

We also perform linear stability analysis on the paramagnetic solution of the
SGBP equations for the 3D EA model (\ref{eq:EAh}).
The stability threshold  $\beta_c$ of the SGBP paramagnetic solution
on the region graph $R_{3D}^{n=2}$
depends on the particular disorder instance. For each
cubic lattice of side length $L$ under the periodic boundary condition,
we generate $32$ independent realisations of the coupling
constants $\{J_{i j}\}$ and then obtain the value of $\beta_c$ for each
disorder realisation. The relationship between the mean value of $\beta_c$
and lattice size $L$ is shown in figure~\ref{fig:3DEA}.
The mean $\beta_c$ value decreases with lattice
side length $L$ and appears to approach the value of
$\beta_c \approx 0.505$ at $L\rightarrow \infty$.
As a comparison, Monte Carlo simulations predicted that
the spin glass phase transition occurs at $\beta_c
\approx 0.893$ (see \cite{Katzgraber-Korner-Young-2006}
and references therein).
We believe that if we enlarge the side length $n$
of the maximal cube region of the used region graph, the critical value
$\beta_c$ at $L=\infty$ will increase and be much more closer to the
value predicted by Monte Carlo simulations.

\section{Conclusion}
\label{sec:conclusion}

In summary, we presented an alternative way of deriving a set of
generalized belief propagation (GBP) equations for a general lattice
statistical physical system. Based on this derivation we proposed a
way of simplifying the GBP equations. 
We also pointed out that, due to the existence of redundance
in the region graph, some of the simplified generalized belief propagation
(SGBP) equations can
be ignored in the numerical iteration process.
We applied the set of SGBP equations to the two-dimensional
and three-dimensional 
Ising model and Edwards-Anderson model.  The numerical results
confirmed that the SGBP message-passing approach significantly outperforms
the conventional BP approach in treating short-range correlations. 
Hopefully this work will stimulate further developments of the
SGBP message-passing approach and the applications of this approach
to finite-dimensional disordered systems and finite-dimensional
optimisation problems.

The three-dimensional Edwards-Anderson model (\ref{eq:EAh}) is in the
spin glass phase at low temperatures, with broken ergodicity. To describe
the low-temperature spin glass property of the EA model,
the effects of ergodicity-breaking need to
explicitly considered in the SGBP equations. This issue
remains to be solved.

\ack

This work was supported by the Knowledge Innovation Program of Chinese
Academy of Sciences (No.~KJCX2-EW-J02), and the National Science
Foundation of China
(grant No.~11121403 and 11225526).

\section*{References}


\providecommand{\newblock}{}

\end{document}